\documentclass[prd,11pt,onecolumn,tightenlines,nofootinbib,groupedaddress]{revtex4}
\usepackage{amsmath,amsfonts,amssymb,amsthm,bbm,psfrag}
\usepackage{subfigure}
\usepackage[colorlinks,citecolor=blue,linkcolor=blue,urlcolor=blue]{hyperref}
\usepackage{mathrsfs}
\usepackage{titlesec} 
\usepackage[T1]{fontenc}
\usepackage{latexsym}
\usepackage{amsbsy}
\usepackage{mathrsfs}
\usepackage{color}
\usepackage{enumerate}
\usepackage{calc}
\usepackage{latexsym}
\usepackage{graphicx,calc,epsfig}

\newcommand{\be}{\begin{equation}}
\newcommand{\ee}{\end{equation}}
\newcommand{\beq}{\begin{eqnarray}}
\newcommand{\eeq}{\end{eqnarray}}
\newcommand{\bea}{\begin{eqnarray}}
\newcommand{\eea}{\end{eqnarray}}

\newcommand{\bra}{\langle}
\newcommand{\ket}{\rangle}

\def\ut#1{\rlap{\lower1ex\hbox{$\sim$}}#1{}}
\newcommand{\ba}{\nopagebreak[3]\begin{eqnarray}}
\newcommand{\ea}{\end{eqnarray}}
\DeclareFontFamily{U}{rsfs}{}         
\DeclareFontShape{U}{rsfs}{m}{n}{<5> rsfs5 <6><7> rsfs7          %
  <8><9><10><10.95><12><14.4><17.28><20.74><24.88> rsfs10}{}     %
\DeclareMathAlphabet{\mathfs}{U}{rsfs}{m}{n}                     %
\newcommand{\mfs}[1]{\mathfs {#1}}                               %

\newcommand{\sH}{{\mfs H}}

\newcommand{\sR}{{\mfs R}}

\newcommand{\sI}{{\mfs I}}
\newcommand{\n}{\nonumber}

\def\pb#1{\rlap{\lower1.5ex\hbox{$\longleftarrow$}}{#1}}
\def\dpb#1{\rlap{\lower1.5ex\hbox{$\Longleftarrow$}}{#1}}
\def\spb#1{\rlap{\lower1.5ex\hbox{$\leftarrow$}}{#1}}
\def\sdpb#1{\rlap{\lower1.5ex\hbox{$\Leftarrow$}}{#1}}

\definecolor{blue}{rgb}{0,0,1}
\definecolor{green}{rgb}{0,1,0}
\definecolor{red}{rgb}{1,0,0}
\definecolor{vio}{rgb}{1,0,1}
\definecolor{ama}{rgb}{1,1,0}

\renewcommand\thesection{\Roman{section}}
\titleformat{\section}{\large\scshape\bfseries\centering}{\thesection.}{.7em}{}
\titleformat{\subsection}{\scshape\bfseries}{\thesubsection.}{.7em}{}
\usepackage{mathtools}

\DeclarePairedDelimiter{\avrg}{\langle}{\rangle}
\DeclarePairedDelimiter{\abs}{\lvert}{\rvert}
\usepackage{braket}

\def\be{\begin{equation}}
\def\ee{\end{equation}}

\def\scrai{\mathscr{I}}
\begin{document}

\title{\large\scshape\bfseries Improved Black Hole Fireworks: Asymmetric Black-Hole-to-White-Hole Tunneling Scenario}

\author{Tommaso De Lorenzo}
\email{tommaso.de-lorenzo@cpt.univ-mrs.fr}
\affiliation{
    Aix Marseille Universit\'e, CNRS, CPT, UMR 7332, 13288 Marseille, and
    Universit\'e de Toulon, CNRS, CPT, UMR 7332, 83957 La Garde, France.
}
\author{Alejandro Perez}
\email{perez@cpt.univ-mrs.fr}
\affiliation{
    Aix Marseille Universit\'e, CNRS, CPT, UMR 7332, 13288 Marseille, and
    Universit\'e de Toulon, CNRS, CPT, UMR 7332, 83957 La Garde, France.
}

\begin{abstract}

A new scenario for gravitational collapse  has been recently proposed by Haggard and Rovelli. Presenting the model under the name of black hole fireworks, they claim that the accumulation of quantum gravitational effects outside the horizon can cause the tunneling of geometry from a black hole to a white hole, allowing a bounce of the collapsing star which can eventually go back to infinity. In this paper we discuss the instabilities of this model and propose a simple minimal modification which eliminates them, as well as other related instabilities discussed in the literature. The new scenario is a time-asymmetric version of the original model with a time scale for the final explosion that is shorter than $m \log m$ in Planck units. Our analysis highlights the importance of irreversibility in gravitational collapse which, in turn, uncovers important issues that cannot be addressed in detail without a full quantum gravity treatment.\end{abstract}

\maketitle

\section{Introduction}

Regular collapse models where the black hole singularity is replaced by some smooth geometry have a long history \cite{bardeen1968non, Frolov:1981mz, Roman:1983zza,Casadio:1998yr, AyonBeato:2000zs,Mazur:2001fv, dymnikova2002cosmological,AshtekarBojowald,hayward2006formation,ModestoLQGBH, Visser:2009pw,Modesto:2010rv,LitimASBH,BambiModestoKerr,Bambi:2013caa,Rovelli:2014cta,Frolov:BHclosed, Mersini-Houghton:2014yq,ModestoNonLocalBH,DeLorenzo:2014pta,Saueressig:2015xua,Frolov:2015bta,DeLorenzo:2015taa}. The {\em leitmotiv} of these models is the attempt to understand issues related to the Hawking information loss paradox
on an effective background spacetime capturing  the idea that black hole singularities must be resolved by quantum gravity effects.
Ideally one would want to justify the relevant physical features of these models in terms of a fundamental quantum theoretical description.
Lacking a precise dynamical description of quantum gravity, their key features are often justified in terms of generic behaviour  that leads to singularity avoidance 
in simplified symmetry reduced models of quantum gravity \cite{Ashtekar:2003hd, Ashtekar:2011ni, Gambini:2013ooa, Gambini:2014qga}.  As one would also expect QFT on curved spacetimes  
to be a valid approximation to quantum dynamics in regions where the gravitational degrees of freedom are well described by a classical background metric of low curvature in Planck units, valuable insights should be accessible through semiclassical methods.
Along these lines a necessary viability criterion for these models is their {\em semiclassical stability}: contributions of quantum fluctuations of a test field in suitable quantum states\footnote{Those satisfying the correct 
boundary conditions that define gravitational collapse.} to the expectation value of the energy momentum tensor must remain small (in Planck units) in semiclassical regions.  
In this article we study the {\em semiclassical stability} of the recently introduced bouncing black hole model proposed in \cite{HalCarloFireworks}.\footnote{A similar scenario in which the same bouncing process happens in much shorter timescales by assuming faster-than-light propagation of a shock-wave from the bounce region is considered in \cite{Barcelo:2014npa,Barcelo:2015noa}.} We find the model to be strongly unstable under small perturbations and consequently we propose a simple but nontrivial modification that avoids these instabilities without modifying the key features of the original idea. 

The paper is organized as follows. In Section \ref{sec:halcarlomodel} we review the definition of the fireworks model.
In Section \ref{scs} we study the semiclassical stability of the fireworks spacetime by computing the expectation value of the energy momentum tensor in a suitable state of a quantum test field on that background.  In order to produce analytic expressions, and thus make clearer our presentation, we will assume that our quantum test field is a massless scalar field and those calculations on the Schwarzschild background will be first done in the approximation where back-reaction is neglected; see Section \ref{next}. We will argue at the end of this section that the result remains valid in the $3+1$ framework where backscattering is taken into account. In Section \ref{corr} we propose a way in which the background of \cite{HalCarloFireworks} could be modified in order to avoid these instabilities as well as other ones described in Section \ref{babacul}. The new model is a time-asymmetric version of the original one, where the black hole phase is followed by an extremely fast explosion with time scale shorter than $m\log m$ in Planck units. Finally, we discuss the implications of such modifications in Section \ref{sw}.

\section{The Fireworks Model}\label{sec:halcarlomodel}
\begin{figure}[t]
\includegraphics[width=0.6\textwidth]{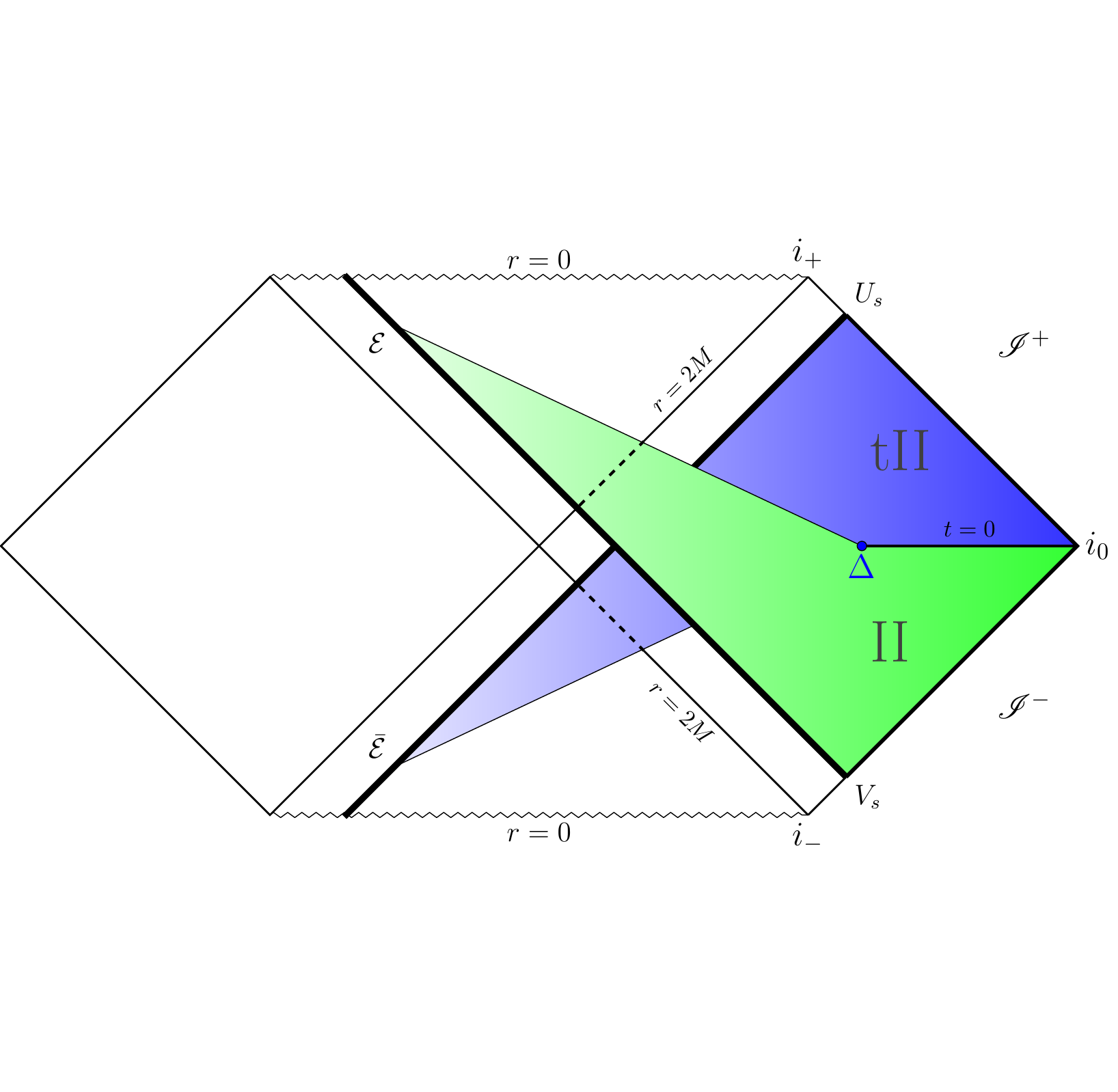} \hfill
\includegraphics[width=0.3\textwidth]{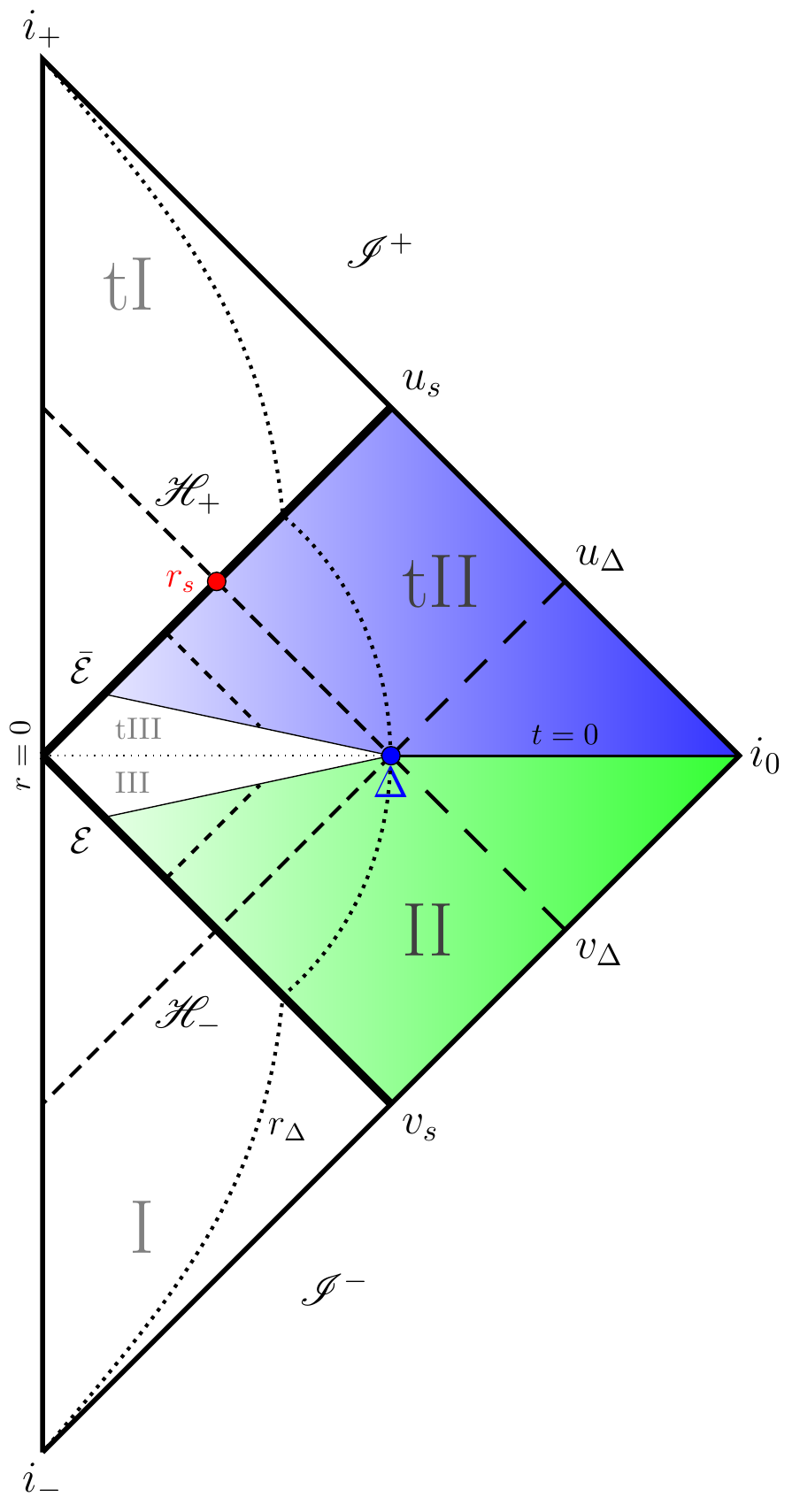}
\caption{Geometry of the black hole fireworks scenario. {\bf Left:} Kruskal-Szekeres diagram, with the two interesting overlapping regions shaded with different colors. {\bf Right:} The resulting completely time-symmetric Carter-Penrose diagram.}
\label{fig:penrose}
\end{figure}

The Penrose diagram of the Haggard-Rovelli \cite{HalCarloFireworks} proposal for a bouncing black hole is shown in Fig.~\ref{fig:penrose}. 
This spacetime corresponds to the collapse of a spherical shell of mass $m$, and it is constructed in terms of patches that are isometric to the Schwalzschild, Minkowski, and an unspecified quantum effective geometry glued together through transition hypersurfaces. In the last region Einstein's equations are not satisfied with any form of classical matter; its presence is interpreted as a modification of the classical dynamics induced by the effect of quantum gravity fluctuations. 

The model can be obtained from the cutting and pasting of regions easily identified in the Penrose diagram of the maximally extended Schwarzschild solution of mass $m$ as follows:  
One first identifies a point $\Delta$ with Kruskal-Szekeres coordinates $(U_\Delta=-V_\Delta, V_\Delta)$ with $V_\Delta>0$ so that $\Delta$ lies in the exterior of the white as well as the black hole regions. One then chooses a null surface $V=V_s$ such that $V_\Delta > V_s$ and a  point $\mathcal{E}$ with coordinates $(U_\mathcal{E},V_\mathcal{E}=V_s)$ and $U_\mathcal{E}>0$, i.e., $\mathcal{E}$ lies on the null surface $V=V_s$ and in the interior of the black hole region. Finally one picks a space-like hypersurface $\Sigma_{ \mathcal{E}\to \Delta}$ connecting $\Delta$ to $\mathcal{E}$ and extends this space-like hypersurface to space-like infinity $i_0$ along the hypersurface $\Sigma_{\Delta\to i_0}$ defined by the condition $t=0$ in Eddington-Finkelstein coordinates. One names Region II the spacetime region bounded by the null surface $V=V_s$ in the past and $\Sigma_{ \mathcal{E}\to \Delta}\cup\Sigma_{\Delta\to i_0}$ in the future. There is a partner Region tII defined in analogy to Region II  by the time reflection $(U,V)\to (-U,-V)$. 
See Fig.~\ref{fig:penrose}-Left. 
The Carter-Penrose diagram of the fireworks model (Fig.~\ref{fig:penrose}-Right) is obtained by inserting the interpolating Regions III+tIII that complete the spacetime to the future of $\Sigma_{\mathcal{E}\to \Delta}$ in Region II up to $\Sigma_{\bar{\mathcal{E}}\to \Delta}$  in Region tII.
The regions $v\le v_s$ and  $u\ge u_s$ are described by Minkowski Region I and Region tI respectively.
The gluing across the null surfaces is done by demanding continuity of the metric; this leads to a distributional energy momentum tensor and the standard interpretation of the null gluing surface as a spherical shell of mass $m$ collapsing to $r=0$ in the past and then bouncing out in the future. The geometry in Region III+tIII is not explicitly defined in the model; however, the absence of singularities require the putative energy-momentum tensor to violate  energy conditions in Region III+tIII. This is interpreted as a spacetime region where quantum gravity effects are large.

The resulting spacetime represents the dynamics of a null in-falling shell of total mass $m$ that bounces at $r=0$ and comes out as a null outgoing shell of the same mass. 
The point $\mathcal{E}$ is the point where the ingoing shell enters (or touches) the quantum Region III, while $\Delta$ is considered as the outmost boundary of the quantum Region III+tIII.  As we will recall below, the time scale of the bounce is argued to be of the order of $m^2$ (in Planck units). This `fast' process makes the dissipation effects of Hawking radiation negligible. This is argued to justify the time-symmetric character of the bouncing scenario.

The spacetime is event-horizon-free, but displays a trapping and an anti-trapping surface. Notice that the past directed outgoing null rays from $\Delta$---defining a null surface that approaches exponentially the trapping surface in the past---represents what we will call the {\em past classicality horizon}, denoted $\sH_{-}$: any observer crossing $\sH_{-}$ will end up falling into the quantum Region III+tIII. More precisely, the domain of dependence $D({\rm III+tIII})$ has a boundary defined by two null surfaces. We call $\sH_{-}$ (resp. $\sH_{+}$) the past (resp. future) null component of that boundary.

To completely specify the model, one has to fix $V_{\Delta}>0$ in order to fix the position of the point $\Delta =(-V_\Delta, V_\Delta)$, and duration of the process which is parametrized by $\Delta V=V_{\Delta}-V_s$.  The condition $V_{\Delta}>0$ implies that the quantum region III+tIII extends outside the Schwarzschild trapping horizon.  This is a central conceptual point in the proposed model: one is allowing large quantum effects to leak out of the Schwarzschild horizon where curvature is low and far from Planckian (here $m\gg1$ in Planck units). In the original paper \cite{HalCarloFireworks}, this is stated by saying that ``there is no reason to trust the classical theory outside the horizon for arbitrarily long times and sufficiently close to $r = 2m$''.  
The authors of \cite{HalCarloFireworks} propose that quantum gravitational effects can be accumulated with ``time'' and become nonnegligible outside the horizon. Accordingly, they  introduced a \emph{nonclassicality parameter} defined along the world-line of a stationary observer sitting at $r=r_\Delta$ for a proper time $\tau$ as
\be\label{qqq}
q=\ell_p^{2-b} \sR\tau^b
\ee
where $\sR$ is a measure of spacetime curvature defined for concreteness in terms of the  Kretschmann invariant $\sR^2=R_{abcd} R^{abcd}=48 m^2/r^6$ and $b$ is a phenomenological parameter of order unity.  For concreteness we take $b=1$ following \cite{HalCarloFireworks}. The parameter $\tau$ is the proper time of the stationary observer from the crossing of the collapsing shell to the point $\Delta$ (see Fig.~\ref{fig:penrose}), that is
\be\label{eq:tau}
\tau = \sqrt{1- \frac{2m}{r_\Delta}} \, \Delta v\;,
\ee
where $v$ is the standard advanced inertial time at $\sI^-$.
The quantity $q$ is maximized for 
\be\label{threes}
\begin{split}
r_\Delta = \frac{7}{3}m.
\end{split}
\ee
 This means that the quantum Region III+tIII extends macroscopically outside the BH horizon.  The bouncing time is defined to be the value of $\Delta v$ for which the 
nonclassicality parameter (linear in $\Delta v$) becomes of order unity. This happens for
\be\label{eq:deltav}
\Delta v\equiv v_\Delta-v_s \sim \tau \sim m^2\;.
\ee
Due to the time symmetry of the construction, the observer at $r_\Delta$ sees the entire bouncing process happening in a proper time $\tau_{\rm tot} = 2 \tau \sim m^2$.
This time scale is very important in what follows and is argued to produce possible experimental observations \cite{Barrau:2014hda,Barrau:2014yka,Barrau:2015uca}.

\section{Semiclassical stability}\label{scs}

The question any classical ansatz spacetime has to be confronted with is whether it admits a physically reasonable quantum state for the test fields living on it.
This requirement represents the first step toward addressing the problem of \emph{back-reaction}. More precisely, in those regions where we can trust the validity of QFT in curved spacetime one expects the quantum dynamics to be well approximated by the semiclassical Einstein's equation  
\be\label{eq:semee}
G_{ab}(g_{ab})=8 \pi\, \avrg{\,T_{ab}(g_{ab})} \;,
\ee
where $\avrg{\,T_{ab}(g_{ab})\,}$ represents the expectation value of the stress-energy tensor of the quantum matter fields propagating on the metric $g_{ab}$. 

The most famous example is the effect of Hawking evaporation on a black hole background \cite{Hawking:1974,hawking1975}. The original computation has been made in the fixed background approximation, completely neglecting the back-reaction. However, this leads to an infinite amount of radiated energy from the hole, clearly in contradiction with energy conservation. Intuitively, one expects the energy radiated to be balanced by a reduction of the Bondi mass of the black hole, leading to the evaporation of the hole and consequently the well-known loss of information paradox \cite{hawking:1976breakdown}.
There are both analytical and numerical works indicating some general features of the evaporation problem \cite{Hajicek19809,Bardeen:1981zz,Parentani:1994ij,Massar:1994iy}; nevertheless, a complete description remains unsolved even in the semiclassical regime of equation \eqref{eq:semee}.

Indeed, the complete backreaction problem could be framed in a formal approximation procedure where one starts by evaluating $\avrg{T_{ab}(g_{ab}^0)}$  on a seed background $g_{ab}^{0}$, and then inserts the result into semiclassical Einstein equations~\eqref{eq:semee} in other to find a new metric $g_{ab}^{1}$: the first-order quantum corrected background metric. Iterating the process one can try to find higher-order corrected line elements eventually converging to a consistent solution $g_{ab}$ of equation \eqref{eq:semee}. Every single step is in general a really difficult task to achieve and the final convergence is not even guaranteed. 

Fortunately, for the present analysis it will be sufficient to solve a much simpler problem. Indeed, the classical initial background $g_{ab}^{0}$---solution of the classical Einstein equations---is a good zeroth approximation of the quantum dynamics only if the quantum corrections coming from $\avrg{T_{ab}(g_{ab}^0)}$  are small in semiclassical regions. This stability of the seed background under the effects of the propagation of quantum test fields living on it will be called {\em quantum-stability} property. In the following of this Section, we will compute $\avrg{T_{ab}(g_{ab}^0)}$ for the model of reference \cite{HalCarloFireworks} and show that it diverges in Region tII. The quantum-stability property, therefore, is not satisfied by the fireworks model.

The computation of $\avrg{\,T_{ab}(g_{ab}^0)}$ on a given unperturbed geometry can be already a very difficult task. In fact, there is in general uncertainties related to the choice of the appropriate physical state for the quantum fields and, at the same time, one needs to appeal to renormalization techniques to eliminate usual $UV$  divergences of QFT in a way that is consistent with general covariance \cite{Wald:1995yp}. Both issues are more subtle and difficult when the background spacetime is not flat. However, the great symmetry of our example and its direct relationship with the well-studied Schwarzschild geometry will allow us to make very precise statements.
 
\subsection{Analytic calculation in the $1+1$ setting}\label{next}
In this section we use spherical symmetry and we neglect back-scattering as well as the influence of modes other than $s$-modes. This allows for an effective description in terms of a $1+1$ theory. These simplifications make possible the analytic computation of effects that qualitatively remain valid in the $3+1$ framework. More precisely, we show that the computation of $\avrg{\,T_{ab}(g_{ab}^0)}$ in the framework of the fireworks background presents a divergent behaviour. Quantum fields are represented by a single massless scalar $\phi$ satisfying the Klein-Gordon equation 
\be\label{uno}
g_{0}^{ab} \nabla_a\nabla_b \phi=0
\ee 
with $g_{0}^{ab}$ the background geometry of the fireworks model in the $r-t$ space. In more detail, the metric in Region II+tII is given by 
\be\label{schw}
ds_{0}^2=-\left(1-\frac{2m}{r}\right) dvdu,
\ee
where $v=t+r_*$ and $u=t-r_*$, with $t$ the Killing parameter and 
\be r_*=r+2m\log\left(\frac{r}{2m}-1\right).\ee
In Region I the metric is 
 \be\label{Mink}
ds_{0}^2=-dvdu_{in},
\ee
where $u_{in}=t_M-r$ and $v=t_M+r$ and $t_M$ is the inertial Minkowski time defined by an observer at the center of the shell. 
The explicit relation between $u_{in}$ and Schwarzschild coordinates can be computed from the matching conditions that follow from
demanding continuity of the metric across the shell, namely
\be\label{doce}
u=u_{\rm in}-4m \log \left( \frac{v_s-u_{\rm in}-4m}{4m}\right).
\ee

\vspace{2ex}\noindent{\scshape\bfseries The state representing gravitational collapse.}\hspace{3ex} The fireworks model describes the physics of a collapsing shell that would classically lead to the formation of a spherical black hole spacetime. This physical situation imposes  clear-cut constraints on the initial conditions of the quantum state of the field $\phi$. On the one hand,
the state for the in-modes of the quantum fields on $\sI^-$ must not be substantially excited. In other words, aside from the zeroth order matter 
distribution defining the collapsing shell that will lead to the formation of the trapped regions in the future, no substantial amount of energy momentum of $\phi$ is poured in from $\sI^{-}$.\footnote{In Appendix \hyperlink{app:HH}{B} we study the contrasting situation where an infinite amount of radiation is sent from infinity: the Hartle-Hawking state.}
This is translated into the demand that the in-modes of the quantum field on  $\sI^{-}$ must be in the vacuum state. A similar boundary condition must hold also for the out-modes in the flat interior of the collapsing shell (Region I). Small perturbations of these conditions could be admitted yet, and this would not change the conclusions that will follow.  

These two conditions are satisfied by the so-called vacuum in-state $\ket{in}$ \cite{Unruh:1976db}, defined as the unique vacuum state of the Fock space where positive frequencies are defined with respect to the mode expansion of solutions of \eqref{uno} of the form
\be
\phi_{in}=e^{i \omega v}, \qquad \phi_{out}=e^{i \omega u_{in}} \;.
\ee
This state corresponds to the required physical condition that there is no incoming radiation from $\sI^-$ as well as no outgoing radiation from inside the shell.
This state represents the idealized physical situation one wants to describe in the context of gravitational collapse.  

\vspace{2ex}\noindent{\scshape\bfseries The region of applicability.}\hspace{3ex} There is uncertainty on the features of the quantum fields in the future domain of dependence of Region III+tIII as the effective $1+1$ geometry is expected not to capture all the physics of the dynamics of the field through that part of the spacetime.  Therefore, all of the components of $\avrg{\,T_{ab}(g_{ab}^0)}$ that we want to compute can be used to describe the energy momentum expectation value only in Region I and in the portion of Region II in the past of $\Sigma_{ \mathcal{E}\to \Delta}$ union the null outgoing ray $u=u_\Delta$ starting at $\Delta$ and reaching $\scrai^+$.

Nevertheless, whatever might be the dynamics in the strong quantum region, we expect to be able to predict without uncertainties at least some of the components of $\avrg{\,T_{ab}(g_{ab}^0)}$ for those points to the future of the horizon $\sH_{+}$. A closer look shows that, due to the decoupling of in and out modes for a conformal theory in the present $1+1$ context, the component $\avrg{\,T_{vv}(g_{ab}^0)}$ is independent of the features of the quantum Region III+tIII. Both $\avrg{\,T_{uu}(g_{ab}^0)}$ and $\avrg{\,T_{uv}(g_{ab}^0)}$, on the other hand, will be modified by quantum gravity effects.
In those regions of applicability, the computation comes out to be a standard computation \cite{PhysRevD.13.2720,birrelldavies:QFTCST}, well illustrated for instance in \cite{fabbri2005modeling}. 

With these preliminary considerations stated, we are now ready to compute the expectation value of the energy momentum tensor in the vacuum in-state defined on the background geometry of the fireworks spacetime. In the region of interest, and for $\avrg{\,T_{vv}(g_{ab}^0)}$ we can simply import the results from the standard calculation on a background given by the gravitational collapse of a shell of mass $m$.   
Following for instance \cite{fabbri2005modeling}, see Appendix \hyperlink{app:HH}{B}, the components of the covariant quantum stress-energy tensor are given by
\be\label{jiji}
\begin{split}
\bra{in}T_{uu}\ket{in} &= \frac{\hbar}{24 \pi}\left[ -\frac{m}{r^3} + \frac{3}{2}\frac{m^2}{r^4} -\frac{8m}{(u_{in}-v_s)^3}-\frac{24 m^2}{(u_{in}-v_s)^4}\right]\\
\bra{in}T_{vv}\ket{in} &= \frac{\hbar}{24 \pi}\left[ -\frac{m}{r^3} + \frac{3}{2}\frac{m^2}{r^4} \right]\\
\bra{in}T_{uv}\ket{in} &= -\frac{\hbar}{24 \pi}\left( 1- \frac{2m}{r} \right) \frac{m}{r^3} \;.
\end{split}
\ee
While the above equations seem to show that $\avrg{\,T_{ab}(g_{ab}^0)}$ is finite everywhere, they do not. The problem is that the Eddington-Finkelstein coordinates used to compute them are not well defined at the trapping horizons: the modes are infinitely oscillating there. A clear analysis of the divergence behavior of the tensor  $\avrg{\,T_{ab}(g_{ab}^0)}$ can be achieved by using good coordinates close to the trapping horizons. The expectation value of the energy momentum tensor in our state can be shown to be regular in whole Region II, see for instance \cite{fabbri2005modeling}. What about Region tII?

Only $\avrg{\,T_{vv}(g_{ab}^0)}$ is relevant for the rest of our analysis: as mentioned above, indeed, it is the only component of the energy momentum tensor for which \eqref{jiji} can be trusted in the future of $\sH_{+}$  independently of the unknown geometry of Region III+tIII. A suitable choice of good coordinates are the Minkowski null coordinates $(u,v_{\rm out})$ in terms of which the metric in Region tI takes the form
\be
ds_0^2=-dudv_{out}.
\ee
 Continuity of the metric across the outgoing shell implies
\be
v=v_{\rm out}-4m \log \left( \frac{u_s-v_{\rm out}-4m}{4m}\right).
\ee
Since, by definition, $\avrg{\,T_{ab}(g_{ab})}$ is covariant, one finds
\be\label{eq:Tkrus}
\bra{in}T_{v_{\rm out}v_{\rm out}}\ket{in} = \left( \frac{dv}{dv_{\rm out}} \right)^2 \bra{in}T_{vv}\ket{in}=\left(\frac{u_s-v_{\rm out}}{u_s-v_{\rm out}-4m} \right)^2 \bra{in}T_{uu}\ket{in}\;.
\ee

In these coordinates and on the outgoing shell, $u_s-v_{\rm out}=2r$. The above quantity  diverges at the white hole trapping horizon $r=2m$ (which in the patchwork construction of \cite{HalCarloFireworks} is close to $\sH_+$) as $(r-2m)^{-2}$. This divergence of $\avrg{T_{ab}}$ is, as we have just shown, explicit in the simplified $1+1$ context.\footnote{In the same way one can show that all the components of the renormalized energy momentum tensor remain finite at the future horizon (close to $\sH_-$).} However, it is a general feature that remains valid in the physical $3+1$ context. Some references where explicit calculations are given are \cite{PhysRevD.15.2088, PhysRevD.21.2185,Balbinot1999301}. All this is implied by the very general result implying that the Hartle-Hawking state is the only globally nonsingular state---satisfying the Hadamard condition that implies the regularity of $\avrg{T_{ab}}$---on the maximally extended Schwarzschild spacetime which is invariant under Killing time translations \cite{Wald:1995yp}. 

We conclude that in the vacuum in-state the expectation value of energy-momentum tensor diverges at the trapping horizon $r=2m$ close to $\sH_+$. However, this horizon is outside the region of validity of our calculation as defined above: it is completely inside the future domain of dependence of the quantum Region III+tIII.\footnote{One can try to interpolate the black hole patch with the white hole one by an effective metric, see for example \cite{Barcelo:2015noa}. This is however not relevant for our discussion.} Nonetheless, the would-be-divergent component is still problematic. The reason is that the trapping horizon and $\sH_{+}$ get exponentially close to each other along the generators of $\sH_{+}$.  

More precisely, let us call   $r_s$ the value of the radius at the intersection of $\sH_+$ and the outgoing shell; see Fig.~\ref{fig:penrose}. 
From the integration of the null geodesic equation, one finds
\be\label{eq:rs}
\begin{split}
r_s &= 2m \left( 1 + W \left[\frac{r_\Delta - 2m}{2m} \exp \left\{\frac{r_\Delta -2m}{2m} - \frac{\Delta u}{4m}\right\} \right] \right) 
\end{split}
\ee 
where $W[x]$ is the Lambert function and $\Delta u = u_s - u_\Delta$.
Clearly, $r_s$ represents the closest point to the past horizon for which we can trust the expression of  $\bra{in}T_{v_{\rm out}v_{\rm out}}\ket{in}$  given in equation eq.~\eqref{eq:Tkrus}. Consequently, it  also gives the largest possible value of that component of the energy momentum tensor. At that point we have
\ba
\n \bra{in}T_{v_{\rm out}v_{\rm out}}\ket{in}\rvert_{r_s} &=& \frac{\hbar}{24\pi} \left( \frac{r_s}{r_s-2m}\right)^2  \left[ -\frac{m}{r_s^3} + \frac{3}{2}\frac{m^2}{r_s^4} \right]\\
& \sim& - \frac{\hbar}{192\pi } \left(\frac{\exp \left \{ \Delta v/(4m) - (r_{\Delta}-2m)/(2m) \right\}}{r_{\Delta}-2m}\right)^2 \;,
\ea
where we used the fact that $r_s \to 2m$ and that, by construction, \be \Delta u = \Delta v .\label{eq:deltas} \ee
Demanding the quantum energy-momentum tensor to be sub-planckian everywhere, we can find a relation between the two parameters of the models, namely $r_{\Delta}$ and $\Delta v$. In fact, ($\hbar = 1$)
\be
\abs{\bra{in}T_{v_{\rm out}v_{\rm out}}\ket{in} } < 1
\ee
implies
\be\label{eq:condition}
\left(\frac{r_{\Delta}}{2m}-1\right) e^{\frac{r_{\Delta}}{2m}-1} > \frac{1}{\sqrt{768 \pi m}\, } e^{\frac{\Delta v}{4m}} \quad \Rightarrow \quad
\frac{r_{\Delta}}{2m}-1 > W\left[ \frac{1}{\sqrt{768 \pi}\, m} e^{\frac{\Delta v}{4m}}\right]\,.
\ee
The longer the lifetime $\Delta v$ of the hole, the more the quantum region must extend out of the classical horizon (as parametrized by $r_{\Delta}$) in order for the stress-energy tensor to be subplanckian along $\sH_+$. In particular, if, as estimated in \cite{HalCarloFireworks}, $r_{\Delta}=\frac{7}{3} m$ (see eq. \eqref{threes}), condition~\eqref{eq:conditiononv} implies
\be\label{eq:bound}
\Delta v \lesssim m \log m\;.
\ee
That is, {if we do not want trans-Planckian behaviors of the renormalized quantum stress-energy tensor, the lifetime of the hole has to be so short that the model would already be ruled out by present observations.  For instance, the characteristic time $\tau=m\log(m)$ would be of about $10$ minutes for the central supermassive black hole in our Milky Way. For the same black hole one could try to tune the parameter $r_{\Delta}$ to allow a lifetime of order $m^2$; however, a simple look at equation \eqref{eq:condition} shows that this would imply extending the quantum region outside of the horizon to include almost the whole of the observable universe. 

\section{Asymmetric Fireworks}\label{corr}

The issues presented in the previous section constrain the white hole lifetime to be much shorter than the one defined in the original paper. 
Similar constraints can be found from simple classical considerations.\footnote{Personal communications with Eugenio Bianchi and Matteo Smerlak.} 
In all cases the problems are related to the instability due to the presence of a white hole horizon: infinite blueshift of perturbations that are well behaved at $\sI^-$. 
Our argument is related to those classical instabilities if we replace the concept of perturbations by quantum fluctuations in the in-vacuum.
However, an important point is that, in all cases, the constraints concern the lifetime of the white hole horizon only. The lifetime of the black hole horizon (which is the one constrained by observations) can be freely set without running into the present type of instabilities. 

This can be easily seen from eq.~\eqref{eq:rs}. The relevant parameter for our discussion is the $\Delta u$ that we identified with $\Delta v$, due to the choice made originally in \cite{HalCarloFireworks} to place the point $\Delta$ on the surface $t=0$. Discarding the identification \eqref{eq:deltas} and following exactly the same procedure, the crucial bound in eq.~\eqref{eq:bound} now becomes
\be\label{eq:condition}
\Delta u \lesssim m \log m\;.
\ee
A possible way out, therefore, is to abandon the time-symmetric nature of the bounce in the original form of the fireworks model.
More precisely, to avoid the time-symmetric condition $\Delta u = \Delta v$ one can modify the construction of the spacetime (Section ~\ref{sec:halcarlomodel}) by choosing the outgoing bouncing shell to come out at a retarded time $U_s$ different from $-V_s$. The resulting spacetime, depicted in Fig.~\ref{fig:a-penrose}, differs from the original one as if the point $\Delta$ has been moved away from the $t=0$ surface along a curve $r=r_{\Delta}$.

\begin{figure}[h]
\includegraphics[width=0.33\textwidth]{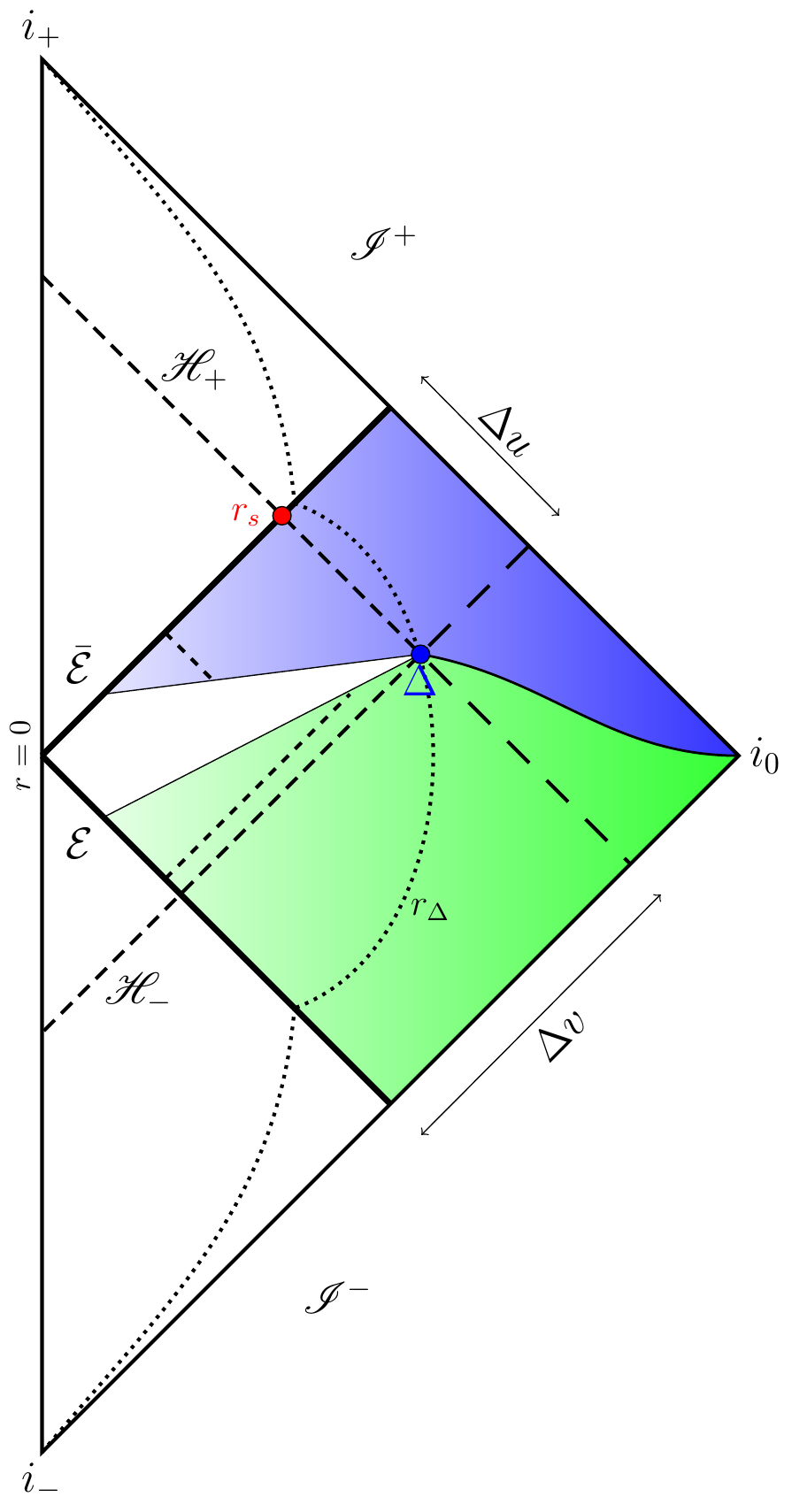}
\caption{Geometry of a asymmetric bouncing scenario. The parameter $\Delta v \sim m^2$ captures both the accumulation time for quantum gravitational effect outside the horizon and the lifetime of the hole. The parameter $\Delta u$, on the other hand, represents the lifetime of the white hole and it is forced by the arguments presented in the text to be of order $m \log m$.}
\label{fig:a-penrose}
\end{figure}

In particular, one can choose the value of $U_s$ such that the quantum stability requirement, expressed by eq.~\eqref{eq:condition}, is satisfied. Moreover, the analysis of the nonclassicality parameter presented at the end of Section~\ref{sec:halcarlomodel} is still precisely valid, and so are eq.~\eqref{threes} and \eqref{eq:deltav}. The accumulation of quantum gravitational effects outside the horizon that allows the black-hole-to-white-hole transition has not been modified, and the above instabilities are removed simply by shortening the lifetime of the white hole horizon.

In Section~\ref{sw} we will largely discuss the nature and the consequences of time asymmetry introduced in our modification of the model. Here we just want to emphasize that the lifetime of the whole process (from collapse to annihilation) remains of the order of $m^2$ as in the original model, much shorter than the $m^3$ time scale predicted by Hawking evaporation.\footnote{In doing this simple comparison between time scales we are making a little abuse of notation. For a more precise statement, see the precise analysis reported in Appendix \hyperlink{app:lifetime}{A}.} This implies that the nature of the time asymmetry is not a dissipative effect due to the Hawking evaporation as one could intuitively expect: the energy radiated after a time of the order of $m^2$ is just of the order of the Planck mass $m_P$. The Hawking effect is negligible and the processes discussed here are basically nondissipative.

\section{Black-hole-to-white-hole instability} \label{babacul}

The modification proposed also removes another related type of instability  studied in \cite{PhysRevLett.33.442,lake1978white,Blau:1989zs,PhysRevD.47.2383,PhysRevD.50.6150}.
The idea is the following. Since a white hole is attractive, any small perturbation of ambient matter will be accelerated toward it. At the same time, since no matter can cross the white hole horizon, after a sufficiently long time, a macroscopic mass will be accreted onto an arbitrarily thin shell close to the horizon and will produce, when interacting with any object coming out from the white hole, a new collapse into a future singularity.
\begin{figure}[t]
\includegraphics[width=0.25\textwidth]{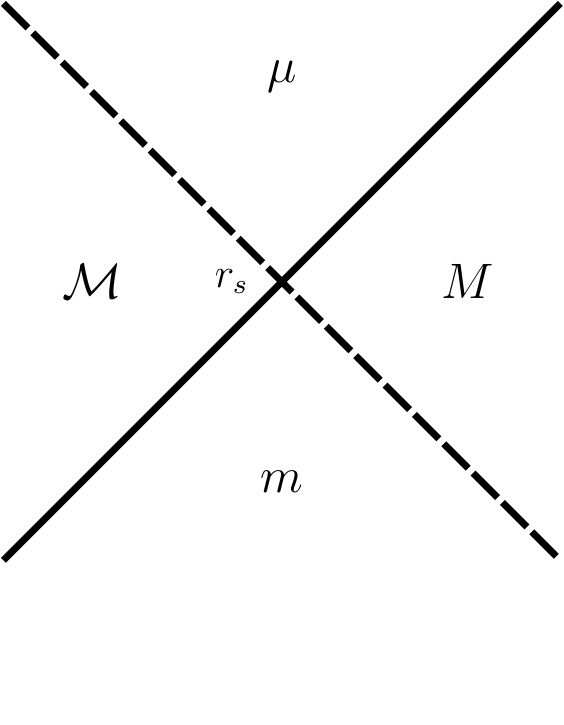} \qquad\qquad\qquad\quad
\includegraphics[width=0.4\textwidth]{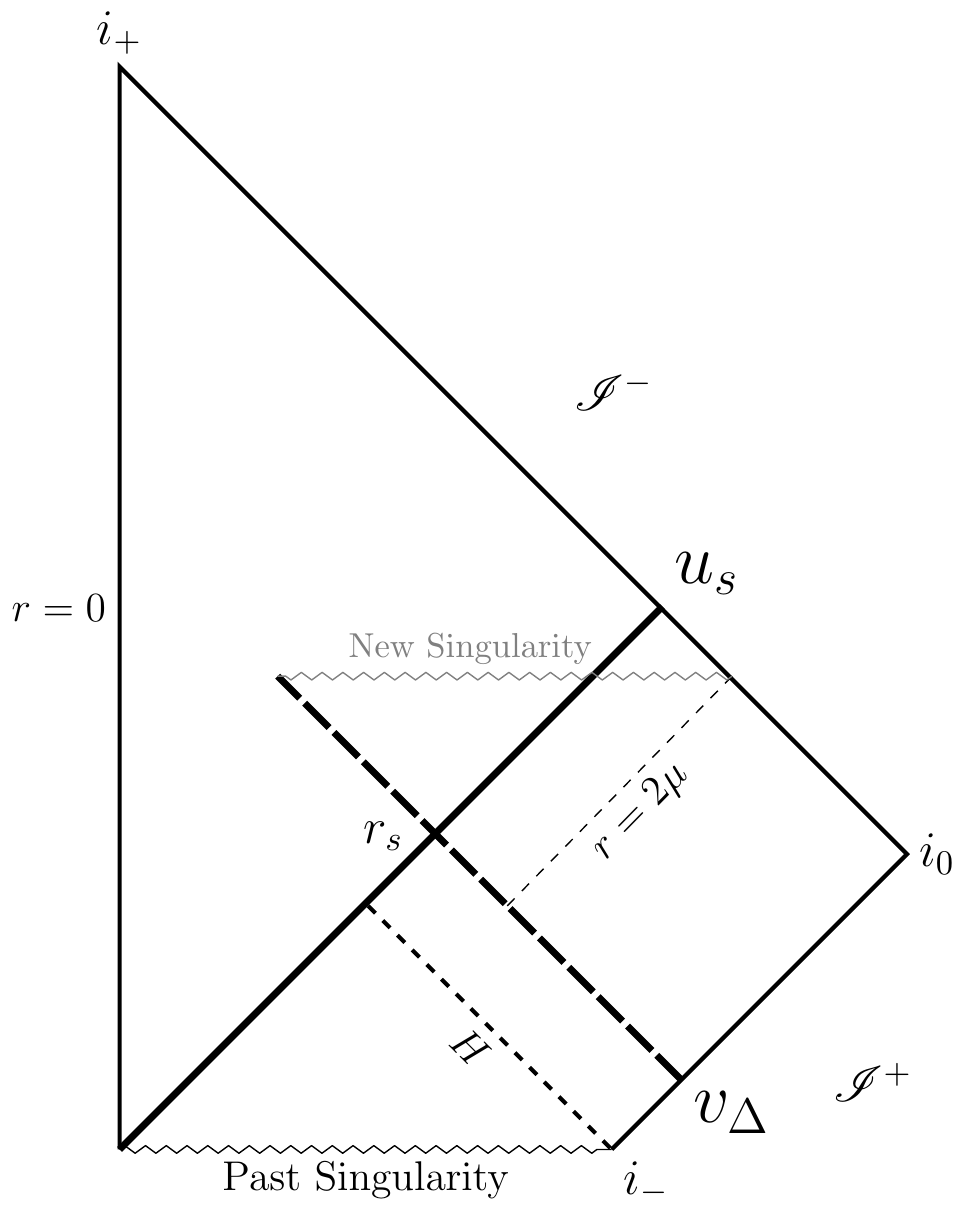}
\caption{
{\bf Left:} The Dray-'t Hooft geometry. Four Schwarzschild patches with different masses are glued together along null shells which intersect at a radius $r_s$. The condition for the geometries to be glued smoothly generates a relation between the four masses and $r_s$. {\bf Right:} The death of a white hole. A white hole emits all its mass $m$ along a massive null shell at the retarded time $u_s$. A small massive perturbation $\nu$ is sent at the advanced time $v_\Delta$ into the white hole geometry and interact with the outgoing shell at $r=r_s$. At the interaction point we have a Dray-'t Hooft geometry with $m=m$, $\mathcal{M}=0$, $M = m+\nu$. The last mass $\mu$ is uniquely determined by the constraint which leads to eq.~\eqref{eq:mu}: $\mu = r_s / (r_s-2m)\nu$. If $r_s$ is lower then $2 \mu$, the emerging shell is captured in the future black hole horizon of the new geometry generated by the interaction and cannot escape to infinity. The white hole is dead recollapsing into a black hole.}
\label{fig:deathWH}
\end{figure}

The interaction between any small matter perturbation of mass $\nu$ sent for instance along the null geodesic $v=v_\Delta$ and the outgoing mass $m$ shell at $r=r_s$ can be described by a Dray-'t Hooft geometry \cite{Dray:1985yt} (Fig.~\ref{fig:deathWH}-Left).  The spacetime for $v>v_\Delta$ and $u>u_s$ is a Schwarzschild geometry with mass $\mu$, given as a function of the initial parameter $m$, $\nu$ and the radius $r_s$ by
\be\label{eq:mu}
\mu = \frac{r_s}{r_s-2m}\,\nu\;.
\ee
It is clear now that if $r_s < 2 \mu$ the outgoing shell is captured inside the new black hole horizon and cannot escape to infinity: any small perturbation $\nu$ interacting with the bouncing shell will cause the system to recollapse into a black hole of mass $\mu$; see Fig.~\ref{fig:deathWH}-Right. Thus, no fireworks can be seen from infinity. The model is, however, still valid if $r_s > 2 \mu$ or equivalently  if
\be
r_s > 2(m+\nu).
\ee
From equation~\eqref{eq:rs} we get
\be
\frac{\Delta u}{4m}  < \left(\frac{r_\Delta-2(m+\nu)}{2m}\right) \log \left(\frac{ {r_\Delta}-2m}{2\nu}\right)\;,
\ee
and assuming $\nu \ll m$ we find again
\be
\Delta u \lesssim m \log \left(\frac{m}{\nu}\right)\;.
\ee
The tunneling process is strongly unstable under perturbations if $\Delta u$, the lifetime of the white hole, is bigger then of order $m \log m$. This argument could surely be discussed together with the other issues that have forced us to consider an asymmetric bouncing scenario, but presented in this way the different time scales involved become clear. This  has been extensively discussed in a recent paper by Barcel\'o et al. \cite{barceloWHinstabilities} in the context of the original symmetric model. The asymmetric modification that we have introduced here also cures this instability. 

\section{Smashing watches}\label{sw}

In this section we want to discuss the physical consequences of the introduction of a time asymmetry in the model. The bouncing process can be described by a quantum field in the $\ket{in}$ vacuum state on $\sI^{-}$ evolving into a final state $\ket{out}$ on $\sI^{+}$. Both states represent an idealized flat initial geometry with an infinitely diluted, but sharply defined, spherical shell carrying mass $m$. More precisely, from the point of view of an observer at infinity, the {\em in} and the {\em out} classical data are just equivalent. 

On the other hand, the semiclassical analysis of the dynamics of the state $\ket{in}$ across the spacetime tells us that the state in the future must be very different from what it was in the past. We have actually shown that the components of $\avrg{\,T_{ab}(g_{ab}^0)}$ are perfectly smooth for $u<u_{\Delta}$ while they are dangerously diverging in some regions to the future $u>u_{\Delta}$.  These divergences can be cured by modifying the background in consistency with this time asymmetry. We have achieved this by shortening the lifetime of the white hole in Section \ref{corr}; see Figure \ref{fig:a-penrose}. 

Nevertheless, in doing so we have preserved the equivalence of the past and future classical data. The point we want to emphasize here is that the time-asymmetric nature of the inner spacetime needed to avoid instabilities should imply strong modifications also in the classical final \emph{out} description of the model, that can be very different from the simple mean field approximation proposed by the fireworks model.

One can illustrate the point in terms of the nonclassicality parameter $q$, eq.~\eqref{qqq}. Recall that the idea is that quantum effects accumulate from $v=v_s$ along the world line of a stationary observer at $r_{\Delta}$ until the quantity $q$ becomes of order one at $v=v_{\Delta}$. This happens after a time of the order of $m^2$. Let us now run the process backward in time. This inverse process is still a bounce now described by an initial state given by $\ket{out}$ evolving into $\ket{in}$. Its dynamics is given by the time reversal of the original one. However, as $q$ only knows about the local geometry, one finds that for the reverse process $q$ is far from unity at $\Delta$. This means that something must be very different for the later observer; something else (not explicitly stated in the model of Figure~\ref{fig:a-penrose}) must contribute to the nonclassicality so that it builds up very much quicker in the inverse process.

If correct, the cause of the shortening of time scales in the future of the bounce must be found in the details of the quantum state of the system beyond the mean-field approximation implicitly used when proposing a background geometry. Notice that the future observer is exposed to quantum gravitational effects coming from the {\em would-be-singularity}---whatever replaces the singularity predicted by the classical theory, i.e. Region III+tIII. These effects must be important enough to drastically reduce the lifetime of the white hole from $m^2$ to $m\log m$. 

But then if these quantum gravitational effects are so strong, why should we trust a semiclassical description at all in the vicinity of the white hole?  
Why should the spacetime become classical again so quickly with the mass $m$ entirely carried by a spherical bouncing shell?
It is hard to address these questions without a full quantum dynamical treatment. 

Nevertheless, the standard collapse process strongly suggests irreversibility already at the classical level. Gravitational collapse is like breaking a watch. This can be intuitively seen, from the classical point of view, by considering the standard spacetime depicting the gravitational collapse of a spherical shell (put the diagram on the right of Figure \ref{fig:deathWH} upright). Initial states given by the shell plus smooth matter and geometry perturbation at $\sI^-$ are special,  they are `low-entropy' states representing our `watches'. They come in different types depending on the details of the initial state. This states are bound to evolve into very complicated final states: smashed watches. This is clear from the fact that only a very precise fine tuning of the features of the state at $\sI^+\cup i_+\cup H$ would evolve backwards to our nice watch at $\sI^-$ (those final states are measure zero in the phase space of possible final states). 

The previous irreversibility mechanism becomes even more apparent if quantum gravity is brought into the discussion. 
Everything that crosses the horizon $\sH_{-}$ will end up at the would-be-singularity exciting degrees of freedom that were not available at low energies.
\begin{figure}[t]
\includegraphics[width=0.33\textwidth]{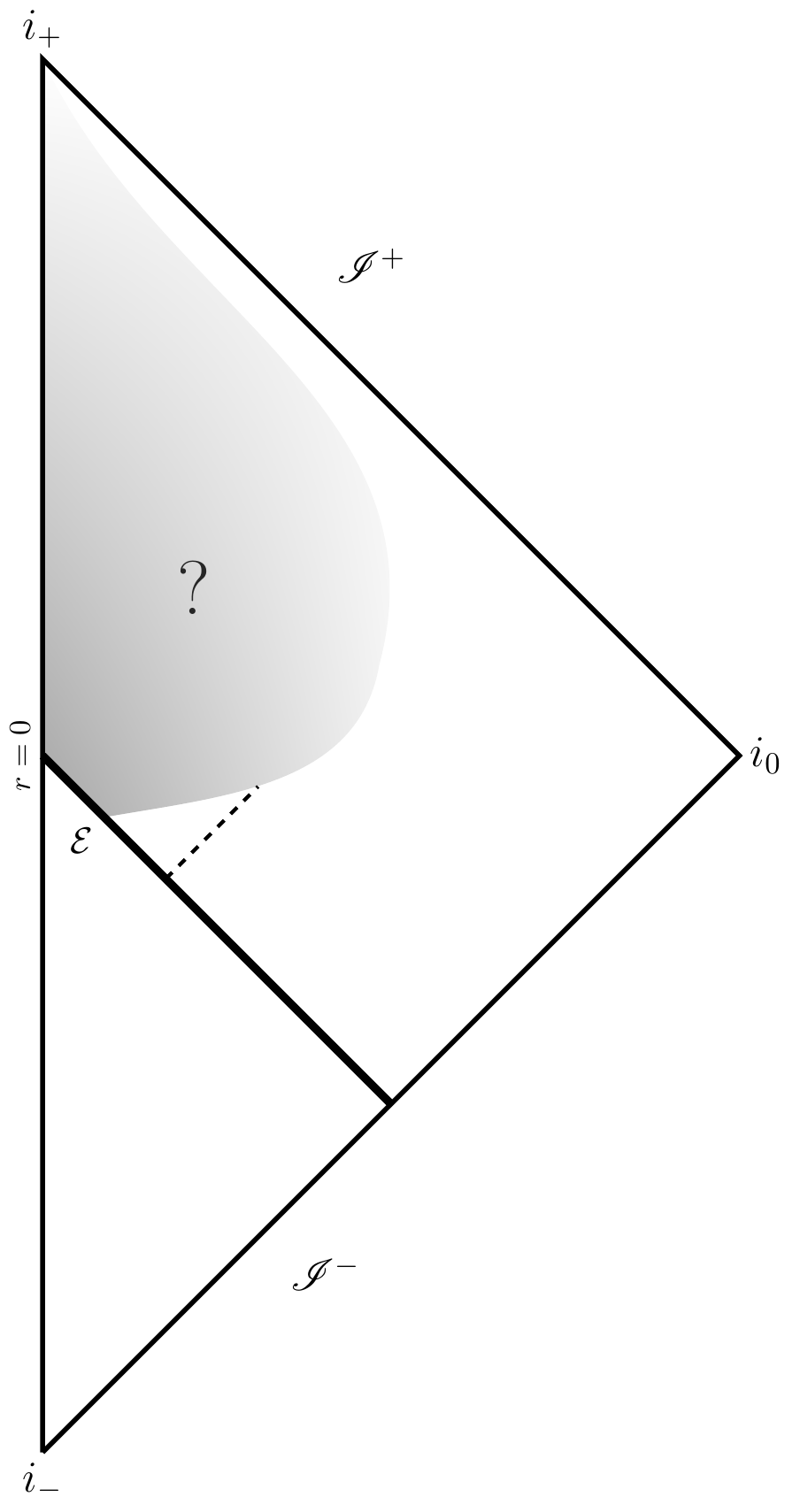}
\caption{Geometry of a general asymmetric scenario. The region where the semiclassical analysis breaks down is shaded. The metric outside is isomorphic to Schwarzschild with mass $m$.}
\label{fig:QuantumFoam}
\end{figure}
The phase space regions available for these falling degrees of freedom can become dramatically larger with the potential effect of further increasing the irreversibility of the overall process. Concretely, as the shell approaches $r=0$ more and more degrees of freedom get excited: from known standard model degrees of freedom (quark-gluon plasma phase, Hagedorn exponential growth of available degrees of freedom, etc.) to beyond standard model degrees of freedom and all the way down to Planck scale. At that ultimate fundamental level, in an approach like LQG, quantum geometries are degenerate: the phase space of available `geometries' at the Planck scale includes a huge number of configurations (microstates) which are simply overlooked in the low energy coarse graining associated with the semiclassical background geometry proposed to describe the process.\footnote{These microstates are responsible for black hole entropy in LQG \cite{G.:2015sda} and have been argued to provide a simple, natural resolution of Hawking's information loss paradox in \cite{Perez:2014xca} in the more conservative framework where Hawking evaporation is the main quantum effect for BHs with $m\gg 1$ \cite{Ashtekar:2005cj}.} All this implies the type of irreversibility proper to systems that satisfy the second law of thermodynamics.

In view of all this we find no reason to discard scenarios where the spacetime does not become semiclassical so quickly to the future of the bounce and where the initial mass $m$ shell dissolves into a quantum substance after the bounce. The details can only be described in the context of full quantum gravity. This very uncertain state of affairs is represented in Figure \ref{fig:QuantumFoam}. 

\section{Conclusions}\label{diss}

We have explored certain instabilities of the fireworks scenario proposed in \cite{HalCarloFireworks} and have proposed a simple way to resolve them.
These instabilities are all associated with the presence of a white hole trapping horizon that is sufficiently long lived.
General considerations demand the gravitational collapse (even in the fast scenario of fireworks where Hawking radiation does not play an important dynamical role) to be time-asymmetric and it  is precisely by allowing such asymmetry that the instabilities are resolved. In this way the black hole phase lasts a time of order $m^2$ followed by an extremely fast explosion where the mass $m$ is radiated back to infinity in a time shorter than $m\log(m)$ in Planck units ($10^{-4} s$ for a solar mass BH, $10^{-9}s $ for a lunar mass BH). The same considerations of the irreversible nature of the gravitational collapse lead to uncertainties in the description of the details of this late bounce. A more precise (not yet available) quantum gravity description of the dynamics across the would-be-singularity could shed light on these details.
It is possible that, despite these uncertainties, the scenarios discussed here could
lead to some generic observable phenomenology (for instance the $m\log(m)$ explosion scale). We leave this question to the experts. 

\section*{Acknowledgment}

We would like to thank H. Haggard, C. Rovelli and T. Josset for discussions. This work has been supported in part by OCEVU Labex (ANR-11-LABX-0060) and the A*MIDEX project (ANR-11-IDEX-0001-02) funded by the "Investissements d'Avenir" French government program managed by the ANR.

\hypertarget{app:lifetime}{\section*{Appendix A: Lifetimes}}
In this Appendix we compare natural time scales that appear in the collapse models. As recalled in Section~\ref{sec:halcarlomodel}, there is the time scale,  introduced in \cite{HalCarloFireworks},  defined as the proper time $\tau$ that an observer seating just outside the horizon at $r_{\Delta}$ has to wait in order for allow quantum gravitational effects to pile up until $q=1$ in that region. This time scale, of order $m^2$, is referred to as the lifetime of the black hole in  \cite{HalCarloFireworks}. However,  when one talks about \emph{lifetime} in black hole physics,  one would rather refer to the retarded time elapsed at $\sI^{+}$ between an initial $u_0$ (roughly defined by detection of the first Hawking quantum), and the complete evaporation of the hole $u_s$ (in our case). 
More precisely $u_0$ can be identified with the retarded time at which the entanglement entropy at $\sI^+$ starts departing significantly from zero. The results of \cite{Bianchi:2014bma} show that this happens for the retarded time corresponding to the collapsing shell shrinking to $r=3m$. We can write  
\be\label{eq:life}
\tau_{\rm life} = u_s - u_0 =  \Delta u+(u_{\Delta}-u_0).
\ee 
The second term can be calculated from the diagrams (the result is the same in different models); the result is
\be
\tau_{\rm life}=\Delta u+\Delta v+\frac{4}{3} m+4m \log(3).
\ee
This means that, to leading scaling order, the lifetime defined in this way coincides with the one used in \cite{HalCarloFireworks}. It is driven by $\Delta v$ when it is chosen to scale with $m$ faster than linearly. In the present models we have
$\tau_{\rm life} \sim \tau \sim m^2$ if $\Delta v\sim m^2$.

\hypertarget{app:HH}{\section*{Appendix B: The Hartle-Hawking state}}

In this Appendix we recall the basic formulae (used in the main text) that allow to compute the renormalized expectation value of the energy momentum tensor in $1+1$ dimensions. Moreover, we compute for completeness the analog of the Hartle-Hawking quantum state in the fireworks background. This state leads to a regular expectation value of the energy momentum tensor in the semiclassical part of the spacetime. It has the well-known thermal properties outside the collapsing shell. However, this state does not represent the physics of gravitational collapse as it does not satisfies the vacuum boundary conditions neither at  $\sI^{-}$ nor inside the collapsing shell as the following calculation shows.

To do this, let us first recall some basic relations \cite{fabbri2005modeling}. Any $1+1$ spacetime is conformally flat and can therefore be written as
\be
ds^2 = - e^{2\rho}dx_+dx_-
\ee
for some function $\rho$ and a double null coordinate system $x_\pm$. The mean value of the covariant stress-energy tensor on some state $\ket{\Psi}$ can be defined to be
\be\label{eq:mink}
\bra{\Psi}T_{\pm\pm}\ket{\Psi}= -\frac{\hbar}{12\pi}\big((\partial_\pm \rho)^2-\partial^2_\pm \rho\big)+\bra{\Psi}:T_{\pm\pm}:\ket{\Psi}
\ee
where $:T_{\pm\pm}:$ is the normal ordered stress-energy tensor. The off-diagonal term is given by
\be\label{eq:minky}
\bra{\Psi}T_{+-}\ket{\Psi}= -\frac{\hbar}{12\pi}\partial_+\partial_- \rho.
\ee

While $\bra{\Psi}T_{\mu\nu}\ket{\Psi}$ is covariant under a coordinate transformation $x_\pm \to \xi_\pm$, the normal ordered stress tensor transforms as
\be
\bra{\Psi}:T_{\xi_\pm \xi_\pm}:\ket{\Psi} =\bra{\Psi}:T_{x_\pm x_\pm}:\ket{\Psi} - \frac{\hbar}{24\pi}\big\{x_\pm,\xi_\pm\big\}
\ee
where
\be
\big\{x_\pm,\xi_\pm \big\} = \frac{d^3 x_\pm/ d\xi_\pm^3}{dx_\pm/d\xi_\pm}-\frac{3}{2}\left(\frac{d^2x_\pm/d\xi_\pm^2}{dx_\pm/d\xi_\pm}\right)^2
\ee
is the Schwarzian derivative.

The terms that are independent of the state $\ket{\Psi}$ are vacuum polarization contributions 
stemming from the conformal anomaly. For example, by identifying $x_+\to v$ and $x_-\to u$ in the Schwarzschild region 
with metric (\ref{schw}), they become:
\ba\label{emi}
\begin{split}
 -\frac{\hbar}{12\pi}\big((\partial_\pm \rho)^2-\partial^2_\pm \rho\big)&=\frac{\hbar}{24 \pi}\left[ -\frac{m}{r^3} + \frac{3}{2}\frac{m^2}{r^4} \right] \n \\
-\frac{\hbar}{12\pi}\partial_+\partial_- \rho&=-\frac{\hbar}{24 \pi}\left( 1- \frac{2m}{r} \right) \frac{m}{r^3}.
\end{split}
\ea 

\vspace{2ex}\noindent{\scshape\bfseries The in-state}\hspace{3ex}

The $\ket{in}$ state is defined with respect to the mode expansion in terms of 
\be
\phi_{in}=e^{i \omega v}, \qquad \phi_{out}=e^{i \omega u_{in}} \;.
\ee
Inside the collapsing shell this state coincides with the Minkowski vacuum: the vacuum polarization vanishes and the normal ordered contribution vanishes.
Outside the collapsing shell we have  
\be
\begin{split}
\bra{in}T_{uu}\ket{in} &= \frac{\hbar}{24 \pi}\left[ -\frac{m}{r^3} + \frac{3}{2}\frac{m^2}{r^4} \right]- \frac{\hbar}{24\pi}\big\{u_{in},u\big\}\\
\bra{in}T_{vv}\ket{in} &= \frac{\hbar}{24 \pi}\left[ -\frac{m}{r^3} + \frac{3}{2}\frac{m^2}{r^4} \right]\\
\bra{in}T_{uv}\ket{in} &= -\frac{\hbar}{24 \pi}\left( 1- \frac{2m}{r} \right) \frac{m}{r^3} \;,
\end{split}
\ee
where we have explicitly written the vacuum polarization terms (\ref{emi}). Using equation (\ref{doce}) one can compute the Schwarzian derivative
term and obtain (\ref{jiji}). 

\vspace{2ex}\noindent{\scshape\bfseries The Hartle-Hawking-like state}\hspace{3ex}

Take the vacuum state $\ket{H}$ of the Fock space where positive frequencies are defined with respect to the mode expansion of solutions of \eqref{uno} of the form
\be
\phi_{in}=e^{i \omega V}, \qquad \phi_{out}=e^{i \omega U} \;,
\ee
where $U$ and $V$ are the Kruskal coordinates for the black hole geometry.
We compute are the  components of the covariant stress-energy tensor of this state 
in the Minkowski patch of the spacetime defining the inside of the collapsing shell (at least the one connected with the Schwarzschild one without touching the quantum region) which is described by the metric
\be
ds^2 = - dv du_{\rm in}\;.
\ee

\vspace{2ex}\noindent{\scshape\bfseries Outside the collapsing shell.}\hspace{3ex}
Outside the collapsing shell and all over its classical chronological future one has
\be
\begin{split}
\bra{H}T_{uu}\ket{H} 
&= \bra{H}T_{vv}\ket{H}=\frac{\hbar}{768\pi m^2}\left(1-\frac{2m}{r}\right)^2\left[1+\frac{4m}{r}+\frac{12m^2}{r^2}\right]\\
\bra{H}T_{uv}\ket{H}&=-\frac{\hbar}{24 \pi}\left(1-\frac{2m}{r}\right)\frac{m}{r^3}
\end{split}
\ee
Notice that these are well behaved in regular coordinates at the past horizon; see \eqref{eq:Tkrus}.
At large $r\to \infty$ we recover the energy momentum tensor of a thermal bath
 \be
\begin{split}
\bra{H}T_{uu}\ket{H} 
&= \bra{H}T_{vv}\ket{H}=\frac{\hbar}{768\pi m^2}\\
\bra{H}T_{uv}\ket{H}&=0.\end{split}
\label{lista10}
\ee

\vspace{2ex}\noindent{\scshape\bfseries Inside the collapsing shell.}\hspace{3ex}
In the Minkowski patch of the spacetime the first term on the right-hand side of eq.~\eqref{eq:mink} is zero and, moreover, by definition the state $\ket{H}$ is such that $\bra{H}:T_{UU}:\ket{H}=\bra{H}:T_{VV}:\ket{H}=\bra{H}:T_{UV}:\ket{H}=0$. Therefore we find
\be
\begin{split}
\bra{H}T_{u_{\rm in}u_{\rm in}}\ket{H} &= -\frac{\hbar}{24\pi}\big\{U,u_{\rm in}\big\}\\
&= \frac{\hbar}{768\pi m^2}\left(1-\frac{8m}{(u_{\rm in}-v_s)}+\frac{48m^2}{(u_{\rm in}-v_s)^2}\right)\\
\bra{H}T_{vv}\ket{H} &= -\frac{\hbar}{24\pi}\big\{V,v\big\}\\
&= \frac{\hbar}{768 \pi m^2}\\
\bra{H}T_{u_{\rm in}v}\ket{H}&=0 
\end{split}
\ee
where we used the matching conditions
\be
\begin{split}
u=-4m \log \left(-\frac{U}{4m}\right)&=u_{\rm in}-4m \log \left( \frac{v_s-u_{\rm in}-4m}{4m}\right)\\
V&=4m \exp\left(\frac{v}{4m}\right)\;.
\end{split}
\ee
For large $r\to \infty$ we recover the thermal fluid in \eqref{lista10}. The collapsing shell in this state is initially 
filled up with radiation at hawking temperature. Due to the contraction of the shell one gets a divergence
of the energy momentum tensor when the shell crosses the origin at $u_{in}=v_s$.

\bibliographystyle{JHEPs}
\bibliography{references}

\end{document}